# Information Content of Hierarchical n-Point Polytope Functions for Quantifying and Reconstructing Disordered Systems

Pei-En Chen[1], Wenxiang Xu[2, 3, *], Yi Ren[1,*], and Yang Jiao[2, 4, *]

[1] Mechanical Engineering, Arizona State University, Tempe AZ 85287, USA

[2] Materials Science and Engineering, Arizona State University, Tempe AZ 85287, USA

[3] College of Mechanics and Materials, Hohai University, Nanjing 211100, P.R. China

[2] Materials Science and Engineering, Arizona State University, Tempe AZ 85287, USA

[*] Corresponding author, E-mail: xwxfat@gmail.com (W.X.), yang.jiao.2@asu.edu (Y. J.) and yiren@asu.edu (Y. R.)

**Abstract**

Disordered systems are ubiquitous in physical, biological and material sciences. Examples include liquid and glassy states of condensed matter, colloids, granular materials, porous media, composites, alloys, packings of cells in avian retina and tumor spheroids, to name but a few. A comprehensive understanding of such disordered systems requires, as the first step, systematic quantification, modeling and representation of the underlying complex configurations and microstructure, which is generally very challenging to achieve. Recently, we introduce a set of hierarchical statistical microstructural descriptors, i.e., the "*n*-point polytope functions" $P_n$, which are derived from the standard *n*-point correlation functions $S_n$, and successively include higher-order n-point statistics of the morphological features of interest in a concise, explainable, and expressive manner. Here we investigate the information content of the $P_n$ functions via optimization-based realization rendering. This is achieved by successively incorporating higher order $P_n$ functions up to $n = 8$ and quantitatively assessing the accuracy of the reconstructed systems via un-constrained statistical morphological descriptors (e.g., the lineal-path function). We examine a wide spectrum of representative random systems with distinct geometrical and topological features. We find that generally, successively incorporating higher order $P_n$ functions, and thus, the higher-order morphological information encoded in these descriptors, leads to superior accuracy of the reconstructions. However, incorporating more $P_n$ functions into the reconstruction also significantly increases the complexity and roughness of the associated energy landscape for the underlying stochastic optimization, making it difficult to convergence numerically.

## 1. Introduction

Disordered systems such as liquid and glassy states of condensed matter [1, 2], colloids [3], granular materials [4, 5], porous media [6, 7], composites [8-10], alloys [11], biopolymer networks [12-15], packings of cells in avian retina [16] and tumor spheroids [17, 18], are ubiquitous in physical, biological and material sciences. The recently discovered disordered low-dimensional quantum materials [19] have demonstrated great potentials for a wide spectrum of device applications. A comprehensive understanding of such disordered systems requires, as the first step, systematic quantification, modeling and representation of the underlying complex configurations and microstructure, which is generally very challenging to achieve. One challenge involves the *in situ* non-destructive characterization of the 3D configuration or microstructure containing key features of interest for these systems on multiple length scales. This challenge has been partially addressed by the development and successful application of advanced non-destructive *in situ* imaging techniques, such as x-ray micro-computed tomography (μCT) [20, 21]. The second challenge involves the development of efficient mathematical frameworks and computational tools for quantitative representation, modeling and reconstruction of such complex disordered systems.

In general, it is notoriously difficult, if not impossible, to derive a mathematically concise yet complete quantification and representation of a disordered system, due to the intrinsic large number of degrees of freedom required to uniquely specify such a system. For example, to uniquely define a crystal structure, one only needs to specify the basis and a set of translation vectors; while to define the structure of glass, the positions of the atoms are required. Recently, a variety of novel approaches have been developed to address the challenges for quantifying disordered systems. In particular, intensive research activities have been devoted to devising *reduced-dimension statistical* quantification and representation based on either complete 3D or lower dimensional structural data set [9, 22-41], and to devising the associated realization rendering methods based on the reduced-dimension representations [42-69]. Examples of established reduced-dimension representation schemes include random field models, statistical descriptor-based representations, and abstract image-based decompositions obtained via machine-learning, to name but a few. Among the descriptor-based representations, a recently developed framework for Hierarchical Materials Informatics, based on complete 2-point statistics and its lower dimensional projections [70-72], can directly yield accurate estimates of properties of a wide class of engineering materials [59, 73-80] and thus, has been incorporated into various integrated computational material design frameworks.

One of the most widely-used descriptor-based set of representations includes the standard n-point correlation (or probability) functions $S_n$, which encode the occurrence probability of specific n-point configurations in the system [8, 81-83]. The complete set of $S_n$ with n = 1, 2, 3, … ∞ provides a complete quantitative microstructure representation, and thus, determines the physical properties of the system under consideration. It has been shown that even the lower-order function $S_2$ can be employed to model a wide spectrum of distinct disordered systems, such as heterogeneous materials [84-91]. In general, it is very challenging to utilize $S_n$ with n≥3, for which one needs to enumerate all distinct n-point configurations and efficiently compute and store their probability of occurrence. The resulting statistical data sets are typically comparable or even larger in size than the original number of degrees of freedom for the system. It has been shown that two-point statistics alone might not be sufficient to represent certain complex microstructures [92-95]. An alternative approach is to employ non-standard lower-order correlation functions such as the cluster functions [96, 97] or surface functions [98, 99], which encode partial higher-order n-point statistics. This method is very effective in capturing specific morphological features (e.g., clustering) [52] but is typically computationally extensive.

Recently, we develop a set of hierarchical statistical morphological descriptors, called the "*n-point polytope functions*" $P_n$ [100], which can provide concise, expressive, and explainable quantification and representation of a wide spectrum of disordered systems. In particular, the polytope functions successively include higher-order n-point statistics of the features of interest in the system, and can be directly computed from multi-modal imaging data, including x-ray tomographic radiographs, optical/SEM/TEM micrographs, and EBSD color maps for quantification of different features of interest. In addition, efficient computational tools to directly extract these statistical descriptors from imaging data has been developed, and the utility of functions for quantifying complex heterogeneous materials and microstructural evolution has been successfully demonstrated [100].

In this work, we systematically investigate the information content of the $P_n$ functions via optimization-based realization rendering. This is achieved by successively incorporating higher order $P_n$ functions up to $n = 8$ to generate material realizations corresponding to the specified set of function, and quantitatively assessing the accuracy of the reconstructed systems via un-constrained statistical morphological descriptors (e.g., the lineal-path function). We examine a wide spectrum of representative random systems with distinct geometrical and topological features, including representative crystalline and disordered particle particles, Poisson distribution of particles, microstructures of concrete and interpenetrating metal-ceramic composites. We find that generally, successively incorporating higher order $P_n$ functions, and thus, the higher-order morphological information encoded in these descriptors, leads to superior accuracy of the reconstructions. However, incorporating more $P_n$ functions into the reconstruction also significantly increases the complexity and roughness of the associated energy landscape for the underlying stochastic optimization, making it difficult to convergence numerically. These results suggest that the $P_n$ functions can provide a superior quantification and representation of complex disordered systems.

## 2. The n-Point Polytope Functions $P_n$ and Realization Rendering via Stochastic Optimization

### 2.1. Definition of the n-point polytope functions $P_n$

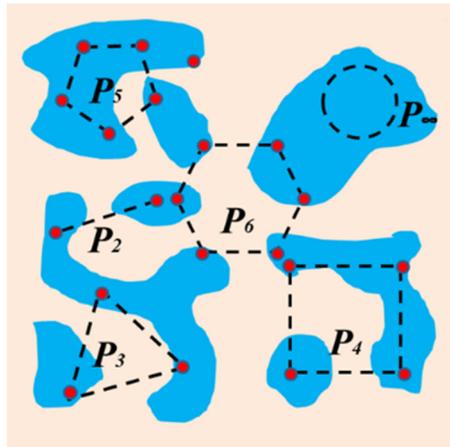

**Fig. 1:** Schematic illustration of stochastic events contributing to the $P_n$ functions in the case of regular polygons.

Without loss of generality, we consider the n-point polytope functions $P_n$ defined for a heterogeneous material (represented as a continuous random field) in which the different structural features are segmented and grouped into different "phases". A simple example is a composite microstructure contains a "matrix phase" and a "particle phase", see Fig. 1. Note that a many-particle system can be easily mapped to a binary random field, e.g., by placing a solid sphere at the center of each particle. The definition of $P_n$ is then given as follows:

$P_n(r) \equiv$ *Probability that all of the n vertices of a randomly selected regular n-point polytope with edge length r fall into the phase of interest.*

Two sets of the $P_n$ functions can be derived based on this definition. The first set involves *n-point regular polygons*, for which the vertex (edge) number *n* can take any positive integer values; and in the limit $n \to \infty$, the shape becomes a circle (in this limiting case, the quantity *r* is the radius of the circle, instead of the edge length). We note that the *n-point polygonal functions* can be computed from both 2D slices and full 3D realizations. The other set involves 3D polyhedra whose edges are of the same length. Only a small number of 3D polyhedra satisfy this condition, including the five *Platonic solids* (i.e., the regular polyhedra: tetrahedron, octahedron, dodecahedron, icosahedron, and cube) and the thirteen *Archimedean solids* (i.e., the semi-regular polyhedra) [101]. Importantly, we note it is clear from the definition that the $P_n$ function is a subset of the corresponding full $S_n$ function. In particular, the former only contains statistics associated with a special subset of the n-point configurations corresponding to a regular n-polytope with different sizes; while the latter encodes statistical information of all n-point configurations. In the case of n = 2, the 2-point polytope function $P_2$ is identical to the standard 2-point correlation function $S_2$.

Figure 1 schematically illustrates the stochastic events that contribute to the $P_n$ functions in the case of regular polygons. For $r = 0$, the polygon reduces to a single point, and $P_n(r=0)$ gives the volume fraction $\varphi$ of the phase of interest (i.e., the probability a randomly selected point falling into the phase of interest). For finite $r$ values, $P_n(r)$ provides n-point spatial correlations in the phase (feature) of interest. For very large $r$ values (e.g., $r\rightarrow\infty$), the probabilities of finding the vertices in the phase of interest are almost independent of one another, thus, we have $P_n(r\rightarrow\infty) \approx \varphi^n$, where $\varphi$ is the volume fraction of the phase of interest in the system. These asymptotic behaviors allow us to introduce a convenient re-scaled form of the $P_n$ functions, i.e.,

$$f_n(r) = \left[P_n(r) - \varphi^n\right] / \left[\varphi - \varphi^n\right] \tag{1}$$

with $f_n(r=0) = 1$ and $f_n(r\rightarrow\infty) = 0$. Finally, we note that one can define "cross-correlation" polytope functions, by requiring a subset of the vertices falling into different phases (features) in the microstructure. In this paper, we will mainly focus on the "auto" polytope functions defined with 2D polygons, in which all of the vertices of the polygons fall into the same phase of interest.

## 2.2. Extracting $P_n$ functions from imaging data

The probability-based definition of the $P_n$ functions allow us to easily compute these function from microstructural data, including both 2D images and 3D digital representations of the material microstructure. For example, in order to compute the value of $P_n(r)$ with $r = r^*$, the following procedure is used:

**(i)** A regular n-polytope (e.g., a n-polygon) with edge length $r^*$ is generated;
**(ii)** This n-polygon is then placed in the material microstructure with randomly selected center location and random orientation, for $M$ times (see Fig. 2a);
**(iii)** Each time the polygon is placed in the system, its vertices is checked to see if they all fall into the phase of interest (i.e., a "success" event); and the total number of success event $M_s$ out of a total of $M$ trials is subsequently recorded;

**(iv)** The rate of success $P_n(r = r^*) = M_s/M$ is computed, which is the probability that a randomly selected n-polygon having all its vertices fall into the phase of interest.

This procedure is repeated for different $r$ values to compute the full $P_n(r)$ function.

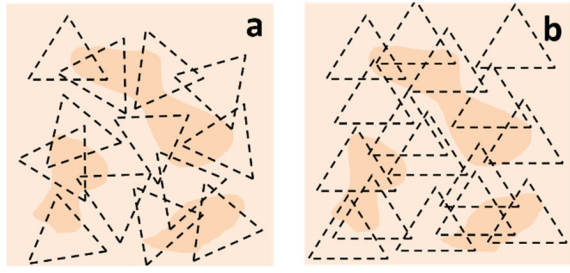

**Fig. 2:** Schematic illustration of different sampling templates for computing $P_n$ functions (in the case of n=3) from images: (a) Isotropic sampling, in which both the location and orientation of the polygon template are randomly selected. (b) Directional sampling, in which only the location of the template is randomly selected.

We note that for a digitized representation of a material microstructure (i.e., an image), a point falls into the phase of interest if it falls into a pixel (or voxel in 3D) of that phase. Therefore, the edge length of the pixel/voxel defines the smallest distance in the system and provides a natural unit for measuring the distance. In addition, besides placing the polytope template with random orientations (i.e., isotropic sampling), one can fix the orientation of the polytopes while placing them at randomly selected locations, see Fig. 2b. We refer to this later case as "directional sampling".

## 2.3. Realization rendering via stochastic optimization

In this section, we briefly describe the Yeong-Torquato (YT) procedure for rendering material realizations via stochastic optimization [102, 103]. In the YT procedure, the reconstruction problem is formulated as an "energy" minimization problem, with the energy functional $E$ defined as follows

$$E = \sum_n \sum_\beta \sum_r \left[ P_n^\beta(r) - \overline{P}_n^\beta(r) \right]^2 \qquad (2)$$

where $\overline{P}_n^\beta(r)$ is a target polytope function of order $n$ along direction $\beta$ and $P_n^\beta(r)$ is the corresponding function associated with a trial realization. In this work, we incorporated $P_n$ functions up to n = 8. The simulated annealing method [104, 105] is usually employed to solve the aforementioned minimization problem. Specifically, starting from an initial trial realization (i.e., *old* microstructure) which contains a fixed number of voxels for each phase consistent with the volume fraction of that phase, two randomly selected voxels associated with different phases are exchanged to generate a *new* trial realization. The associated $P_n$ functions are sampled from the new trial realization and the associated energy is evaluated, which determines whether the new trial realization should be accepted or not via the probability:

$$p_{acc}(old \rightarrow new) = \min\left\{1, \ \exp\left(\frac{E_{old}}{T}\right) \bigg/ \exp\left(\frac{E_{new}}{T}\right)\right\} \quad (3)$$

where $T$ is a virtual temperature that is chosen to be high initially and slowly decreases according to a cooling schedule [48, 51]. The above process is repeated until $E$ is smaller than a prescribed tolerance, which we choose to be $10^{-6}$ here. Generally, several hundred thousand trials need to be made to achieve such a small tolerance, for which we consider convergence is achieved.

An important step in the reconstruction process is to efficiently compute the correlation functions from the trial microstructure, which is especially crucial for the reconstruction of large-scale 3D realizations. The key idea is to only compute the change in the $P_n$ functions due to the switch of randomly selected pixel pair, instead of re-compute the functions from scratch. This can be implemented by identifying the polygonal sampling templates that have been affected by the pixel exchange and recomputing the contributions from these templates after the pixel exchange. We note that when $n$ gets larger, the number of affected $n$-polygonal templates also increases rapidly. Since our focus in this work is to investigate the information content of the hierarchical set of $P_n$ functions using relatively small 2D systems (e.g., ~ 100 by 100 pixels), we afford to directly recompute the $P_n$ functions from new trial realizations without significantly increasing the computational cost.

## 3. Information Content of $P_n$ Functions via Realization Rendering

In this section, we investigate the information content of the $P_n$ functions using a variety of heterogeneous systems, including ordered and disordered particle packings respectively representing crystalline and liquid state of matters, overlapping particles with a Poisson distribution of particle centers, interpreting microstructure of metal-ceramic composite, and concrete microstructure. This is achieved by incorporating additional $P_n$ functions of successively higher orders into the realization reconstructions, starting with the fundamental two-point function $P_2$ (or equivalently $S_2$). The accuracy of the reconstructed realizations are subsequently accessed using statistical descriptors that are not used in the reconstruction, such as the lineal-path function $L(r)$ [52, 106], which provides the probability that a randomly placed line segment of length $r$ entirely lying in the phase of interest. In particular, the lineal-path function $L(r)$ will be computed from both the original system and the reconstructed realization. The accuracy metric $\psi$, defined as the sum of the absolute differences of these two functions over all $r$ values, will also be computed, i.e.,

$$\Psi = \sum_r |L(r) - L^*(r)| \tag{4}$$

where $L(r)$ and $L^*(r)$ are respectively computed from the reconstructed realization and the original system. The level of accuracy of the realizations, quantified via ψ, reflects the statistical information encoded in the $P_n$ functions. For example, higher accuracy indicates essential additional structural information is encoded in the newly added functions. In the following realization rendering cases, the size of the reconstructed system is 80 by 80 pixels without further elaboration, and periodic boundary conditions are employed.

### 3.1 Ordered packing of congruent spheres

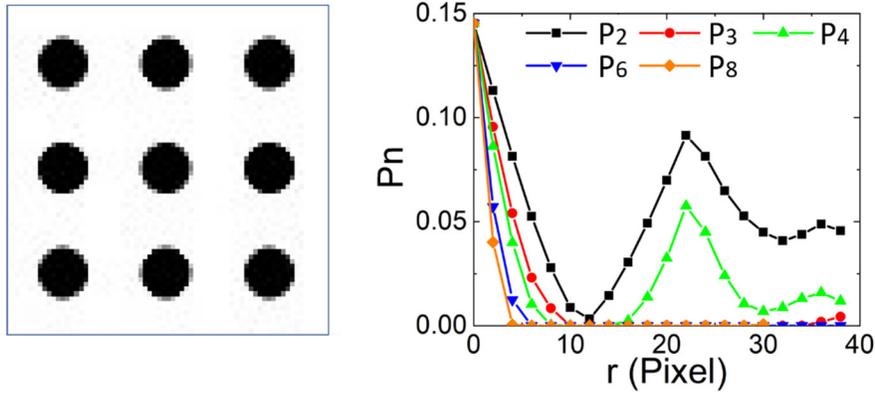

**Fig. 3:** A simple 2D crystalline packing of congruent spheres on a square lattice (left) and the associated $P_n$ function for the particle phase (right).

We begin with a simple 2D crystalline packing of congruent spheres on a square lattice (see Fig. 3a). The volume fraction of the particle phase is phi = 0.14. It can be clearly seen from Fig. 3b that both the $P_2$ (i.e., $S_2$) and $P_4$ functions exhibit strong oscillations, which is a manifestation of the underlying 4-fold symmetry of the structure. The other $P_n$ functions (i.e., n=3, 6, 8) do not exhibit significant oscillation beyond an initial decay, reflecting the fact that the system is composed of compact particles as building blocks, but does not possess strong n-point correlations on large length scales.

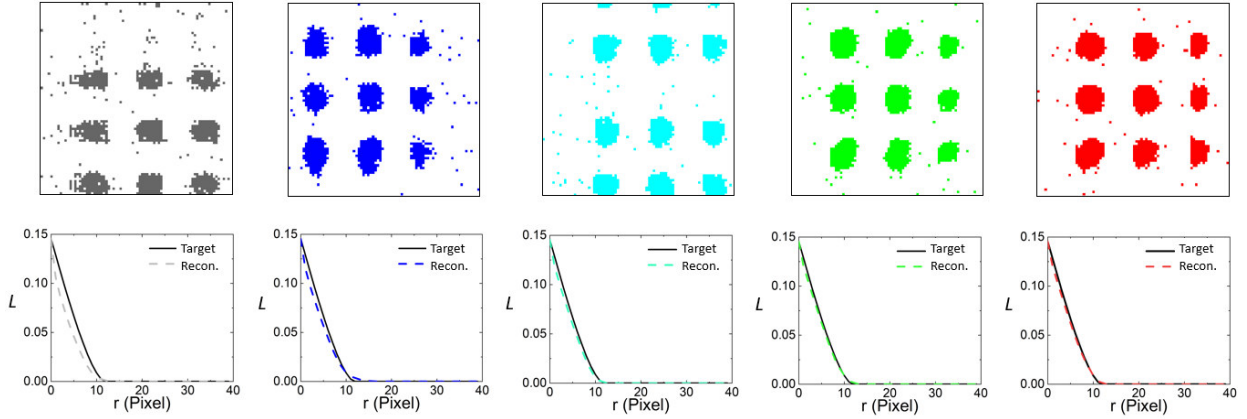

**Fig. 4:** Realizations of 2D crystalline packing of congruent spheres on a square lattice (upper panels) and the associated lineal-path functions (lower panels) obtained via stochastic reconstruction by successively incorporating higher-order $P_n$ functions. The functions incorporated from left to right are respectively: $P_2$, $\{P_2, P_3\}$, $\{P_2, P_3, P_4\}$, $\{P_2, P_3, P_4, P_6\}$, and $\{P_2, P_3, P_4, P_6, P_8\}$.

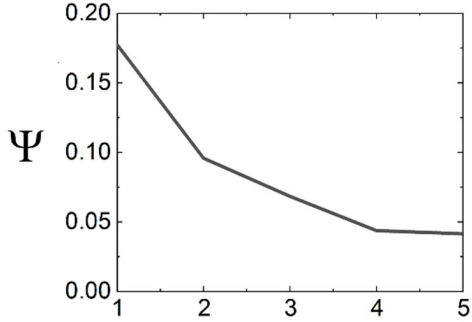

**Fig. 5:** Accuracy metric $\psi$ defined in Eq. (4) associated with reconstructions incorporating different sets of $P_n$ functions. The horizontal axis indicates the number of functions incorporated in the reconstruction.

Figure 4 shows the realizations (upper panels) obtained via stochastic reconstruction by successively incorporating higher-order $P_n$ functions. The functions incorporated from left to right are respectively: $P_2$, $\{P_2, P_3\}$, $\{P_2, P_3, P_4\}$, $\{P_2, P_3, P_4, P_6\}$, and $\{P_2, P_3, P_4, P_6, P_8\}$. It can be seen that the spatial arrangement of the particles on the square lattice is reproduced in all cases. This is because the structural information for the highly ordered 4-fold symmetric arrangement is already captured by the lowest order function $P_2$, which is further reinforced by incorporating $P_4$. Interestingly, the shape of the particle is improved due to incorporation of additional functions, i.e., $P_3$, $P_6$, and $P_8$. This is consistent with information content of these functions, i.e., $P_3$, $P_6$, and $P_8$ functions do not possess long-range oscillations and thus, only encode information on the morphology of the particles. The lineal-path functions $L$ are computed from the reconstructed realizations and compared to that of the original system (see Fig. 4, lower panels). It can be seen as higher order $P_n$ functions are successively incorporated, the $L$ function of the reconstructed realizations matches the original system better. This is consistent with the quantification using accuracy metric shown in Fig. 5.

**3.2 Disordered packing of congruent spheres**

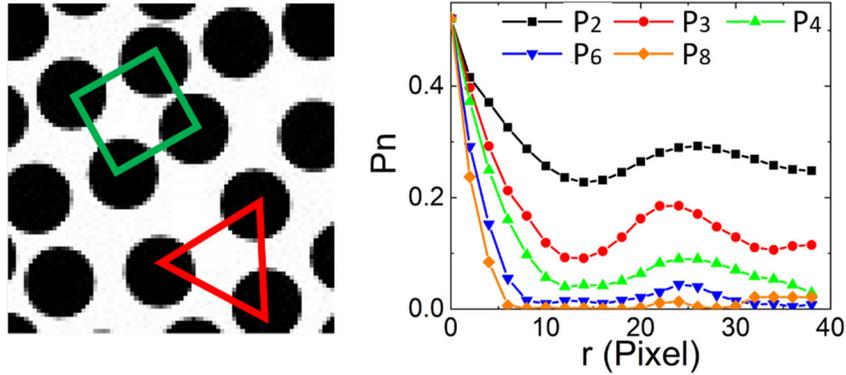

**Fig. 6:** A 2D disordered packing of congruent hard spheres generated via Monte Carlo simulations (left) and the associated $P_n$ function for the particle phase (right).

Fig. 6 shows the quantification of a 2D microstructure composed of equal-sized hard spheres in a matrix [48], i.e., a packing (see Fig. 6, left panel). The sphere packing is generated using Monte Carlo simulations [101]. The $P_n$ functions for the particle phase with n =2, 3, 4, 6 and 8 are shown in Fig. 6 right panel. Similar to the crystalline packing case discussed in Sec. 3.1, all $P_n$ functions initially decay from the volume fraction $\varphi=0.48$ as $r$ increases from 0. The positions of the first minimum in the $P_n$ functions for small n values roughly correspond to the linear size of the particle (~ 12 pixels). After the initial decay, all $P_n$ functions studied here except for $n = 8$ exhibit relatively significant oscillations and the first peaks in different $P_n$ functions occur at approximately the same $r$ values. These oscillations respectively indicate strong pair, triangle, square and hexagonal correlations on different length scales in the system. In Fig. 6, we illustrate examples of such correlations, which are all associated with the mean nearest neighbor separate distance, i.e., the distance associated with the first peak in $P_2$. These correlations result from the tendency for the particles to self-organize at high densities.

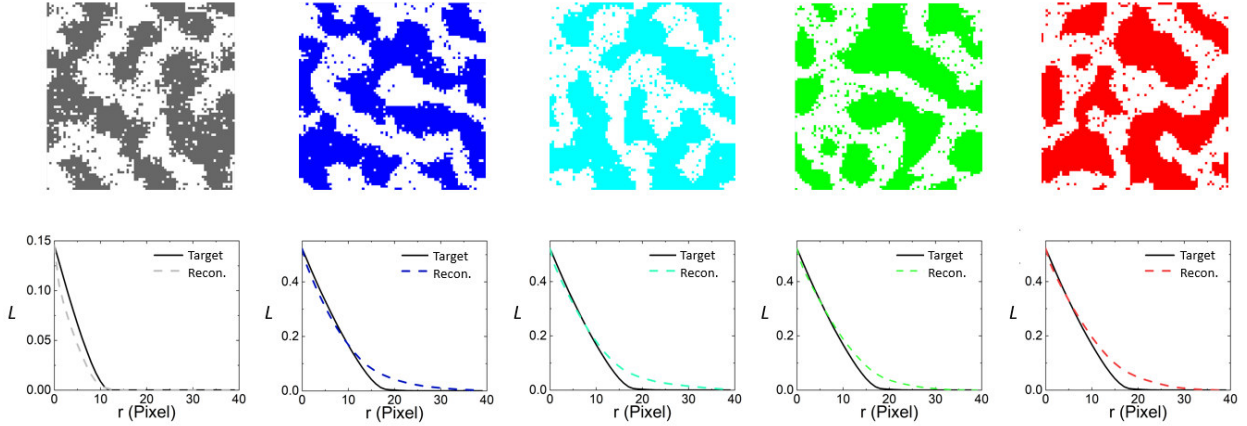

**Fig. 7:** Realizations of disordered packing of congruent hard spheres (upper panels) and the associated lineal-path functions (lower panels) obtained via stochastic reconstruction by successively incorporating higher-order $P_n$ functions. The functions incorporated from left to right are respectively: $P_2$, $\{P_2, P_3\}$, $\{P_2, P_3, P_4\}$, $\{P_2, P_3, P_4, P_6\}$, and $\{P_2, P_3, P_4, P_6, P_8\}$.

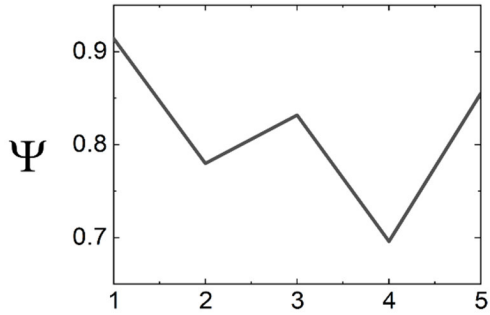

**Fig. 8:** Accuracy metric $\psi$ defined in Eq. (4) associated with reconstructions incorporating different sets of $P_n$ functions. The horizontal axis indicates the number of functions incorporated in the reconstruction.

Figure 7 shows the realizations (upper panels) obtained via stochastic reconstruction by successively incorporating higher-order $P_n$ functions. The functions incorporated from left to right are respectively: $P_2$, $\{P_2, P_3\}$, $\{P_2, P_3, P_4\}$, $\{P_2, P_3, P_4, P_6\}$, and $\{P_2, P_3, P_4, P_6, P_8\}$. It can be seen that in all reconstructions, the connectivity of the particle phase has been significantly overestimated. In particular, instead of reproducing individual compact particles, a single connected phase with a characteristic ligament size comparable to the sphere diameter is produced. The strong oscillation in the correlation functions is realized by the inter-ligament spacing and correlations. This is because the particle volume fraction $\varphi=0.48$ is close to percolation [107] [108]. These results also indicate that the $P_n$ functions (up to n = 8) do not encode topological connectedness information. On the other hand, it has been shown that incorporating cluster functions [41, 52] can lead to significantly improved reconstructions capturing the connectivity information. The lineal-path functions $L$ also are computed from the reconstructed realizations and compared to that of the original system (see Fig. 7 lower panels). Since $L$ is sensitive to clustering, which is overestimated in all realizations, successively incorporating higher-order $P_n$ functions does not lead successively more accurate reconstruction. Instead, the more complex energy landscape associated with higher-order $P_n$ functions might lead to slower convergence of the reconstruction, and higher probability that the system gets stuck in a shallow local minimum. This is consistent with the quantification using accuracy metric shown in Fig. 8.

### 3.3 Overlapping congruent spheres

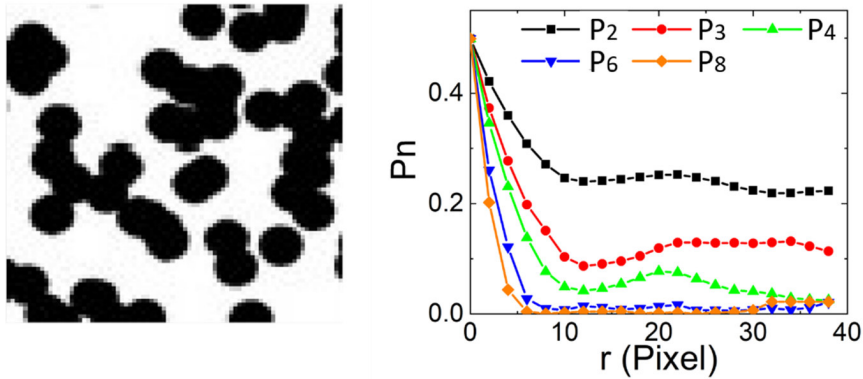

**Fig. 9:** A 2D disordered packing of congruent overlapping spheres with a Poisson distribution of the particle centers (left) and the associated $P_n$ function for the particle phase (right).

Fig. 9 shows the quantification of a 2D microstructure composed of equal-sized overlapping spheres in a matrix [48] (see Fig. 9, left panel). The spheres are randomly placed in the matrix without any built-in spatial correlations. Fig. 9 right panel shows the $P_n$ functions for the particle phase with n =2, 3, 4, 6 and 8. Similar to the previous systems, all $P_n$ functions initially decay from the volume fraction $\varphi$=0.47 as $r$ increases from 0. After the initial decay, the $P_n$ functions are virtually flat, indicating that the particles possess no spatial correlations of any symmetry on any length scales beyond the diameter of the particles. We note that for the totally random overlapping sphere system, the $P_n$ functions possess the analytical expression $P_n(r) = \exp[-\rho v_n(r; R)]$, where $\rho$ is the number density of the spheres in the system (i.e., number of spheres per unit volume) and $v_n(r; R)$ is the volume of the union of $n$ spheres with radius $R$ with centers placed at the vertices of a n-polytope with edge length $r$.

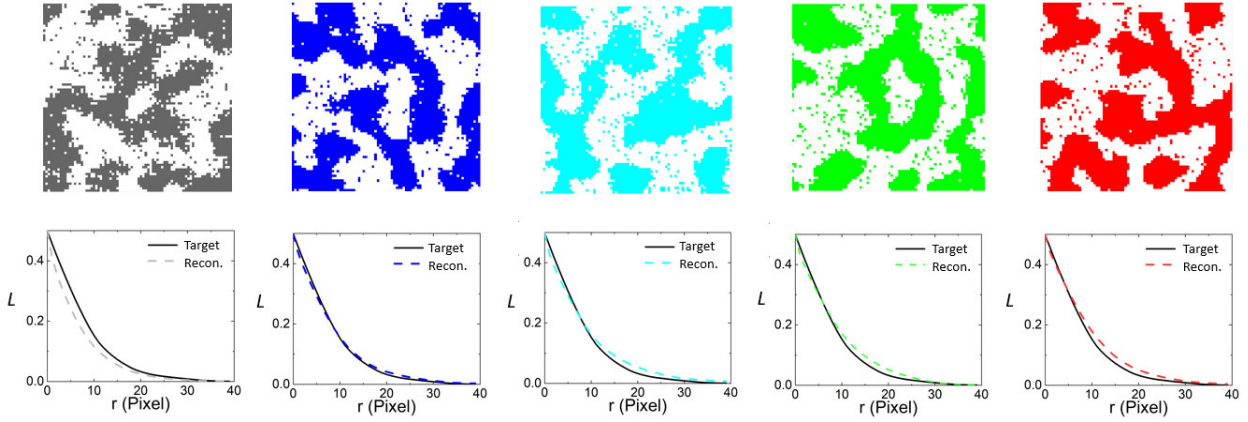

**Fig. 10:** Realizations of overlapping sphere packing (upper panels) and the associated lineal-path functions (lower panels) obtained via stochastic reconstruction by successively incorporating higher-order $P_n$ functions. The functions incorporated from left to right are respectively: $P_2$, $\{P_2, P_3\}$, $\{P_2, P_3, P_4\}$, $\{P_2, P_3, P_4, P_6\}$, and $\{P_2, P_3, P_4, P_6, P_8\}$.

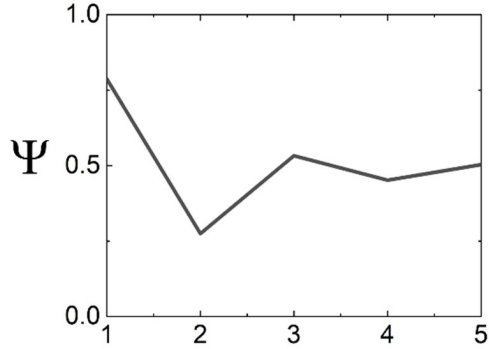

**Fig. 11:** Accuracy metric $\psi$ defined in Eq. (4) associated with reconstructions incorporating different sets of $P_n$ functions. The horizontal axis indicates the number of functions incorporated in the reconstruction.

Fig. 10 shows the realizations (upper panels) obtained via stochastic reconstruction by successively incorporating higher-order $P_n$ functions. The functions incorporated from left to right are respectively: $P_2$, $\{P_2, P_3\}$, $\{P_2, P_3, P_4\}$, $\{P_2, P_3, P_4, P_6\}$, and $\{P_2, P_3, P_4, P_6, P_8\}$. In all the reconstructions, the connectivity of the particle phase has been correctly reproduced, as the overlapping particles in the original system form a single connected phase, in contrast to the hard sphere packing case. Since all the $P_n$ functions contains no essential information beyond the particle diameter (~ 12 pixels), incorporating these functions does not lead to improvement of the reconstruction accuracy for the overlapping sphere system. This is consistent with the quantification using accuracy metric shown in Fig. 11.

### 3.4 Concrete microstructure

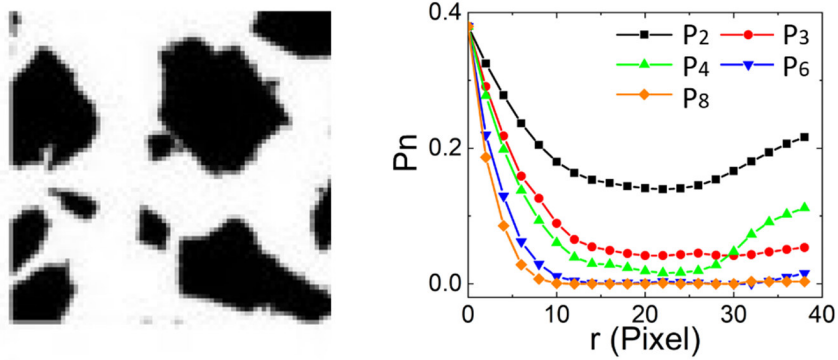

**Fig. 12:** A 2D slice of a concrete microstructure in which the cement paste is shown in white and the rocks are shown in black (left) and the associated $P_n$ function for the rock phase (right).

Fig. 12 shows the quantification of a 2D concrete microstructure composed of reinforcement rocks (shown in black) and the cement paste (shown in white) [52] (see Fig. 12, left panel). The rock particles possess complex polygonal morphologies and a wide size distribution. Fig. 12 right panel shows the $P_n$ functions for the rock phase with n =2, 3, 4, 6 and 8. All $P_n$ functions initially decay from the volume fraction $\varphi$=0.38 as $r$ increases from 0. The $r$ value associated with the first minimum in the functions provides the average particle size in the system, i.e., ~ 20 pixels. After the initial decay, the lower-order $P_n$ functions (e.g., n≤4) increase, reflecting the spatial correlations resulted from the mutual exclusion effects of the rock particles. The correlations are much weaker compared those in hard-sphere systems, mainly due to the anisotropy and size polydispersity of the rock particles.

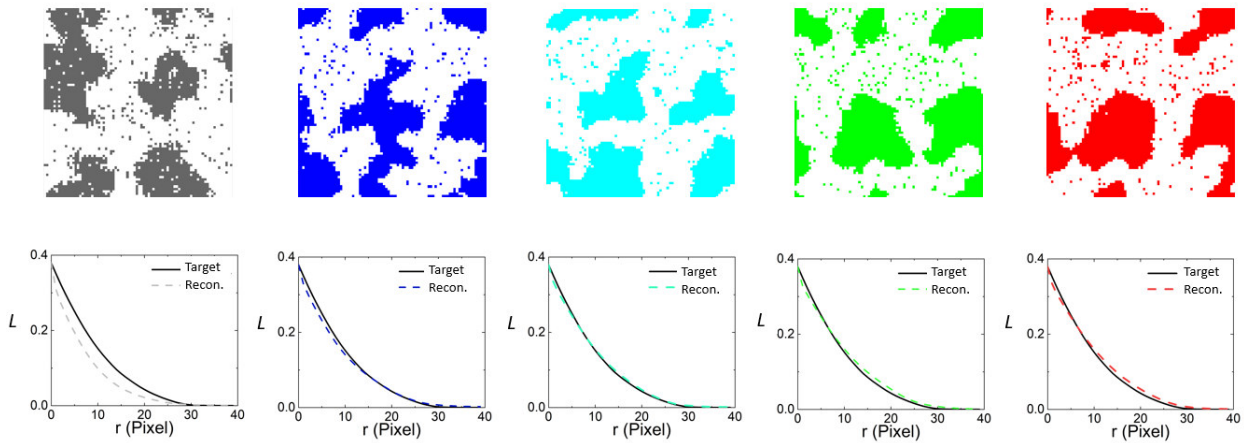

**Fig. 13:** Realizations of concrete microstructures (upper panels) and the associated lineal-path functions (lower panels) obtained via stochastic reconstruction by successively incorporating higher-order $P_n$ functions. The functions incorporated from left to right are respectively: $P_2$, $\{P_2, P_3\}$, $\{P_2, P_3, P_4\}$, $\{P_2, P_3, P_4, P_6\}$, and $\{P_2, P_3, P_4, P_6, P_8\}$.

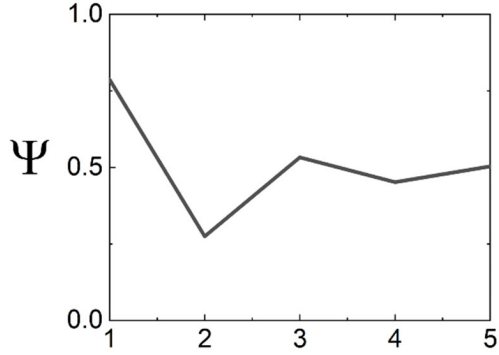

**Fig. 14:** Accuracy metric ψ defined in Eq. (4) associated with reconstructions incorporating different sets of $P_n$ functions. The horizontal axis indicates the number of functions incorporated in the reconstruction.

Fig. 13 shows the realizations (upper panels) obtained via stochastic reconstruction by successively incorporating higher-order $P_n$ functions. The functions incorporated from left to right are respectively: $P_2$, $\{P_2, P_3\}$, $\{P_2, P_3, P_4\}$, $\{P_2, P_3, P_4, P_6\}$, and $\{P_2, P_3, P_4, P_6, P_8\}$. In the reconstructions, individual rock particles can be clearly distinguished. As higher-order $P_n$ functions are successively incorporated, the shape and morphology of the rock particles are improved. This can be seen both from visual inspection of the reconstructed realizations and the quantitative comparison of the lineal-path functions (Fig. 13, right lower panels). In the case that a larger number of $P_n$ functions are incorporated, the complexity of energy landscape increases dramatically, leading to very slow convergence of the reconstruction. This also leads to the slight increase of ψ shown in Fig. 14.

**3.5 Interpenetrating metal-ceramic composite**

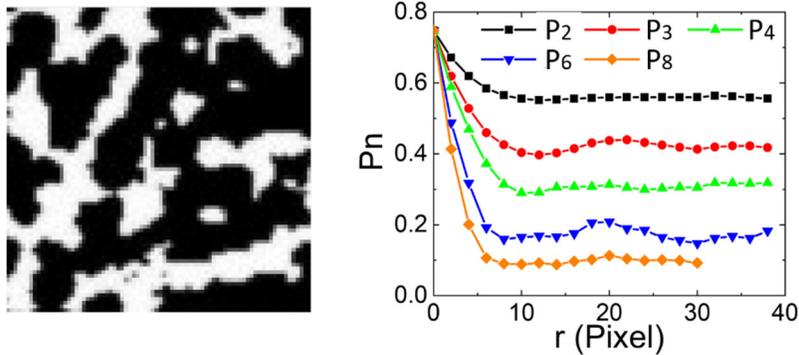

**Fig. 15:** A 2D slice of a interpenetrating microstructure of a metal-ceramic composite composed of the boron-carbide phase (black) and the aluminum phase (white) (left) and the associated $P_n$ function for the boron-carbide (ceramic) phase (right).

Fig. 15 shows the quantification of a 2D an interpenetrating metal-ceramic composite composed of the boron-carbide phase (shown in black) and the aluminum phase (shown in white) [51] (see Fig. 9a). This system contains "ligaments" of similar width instead of "particles", and possesses connected material phase, in contrast to the hard particle packings. Fig. 15 right panel shows the $P_n$ functions for the ceramic phase with n =2, 3, 4, 6 and 8. Similar to the previous systems, all $P_n$ functions initially decay from the volume fraction $\varphi=0.76$ as $r$ increases from 0. The $r$ value associated with the first minimum in the functions provides the average ligament width in the system, i.e., ~ 10 pixels. After the initial decay, the $P_n$ functions exhibit weak oscillations for small and intermediate $r$ values, characterizing the exclusion effects between the ligaments.

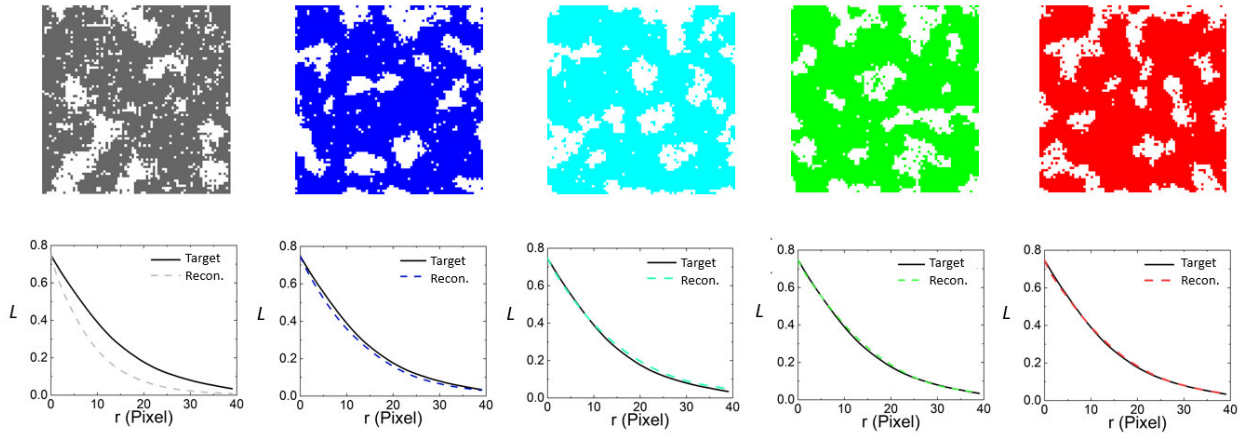

**Fig. 16:** Realizations of the interpenetrating microstructure (upper panels) and the associated lineal-path functions (lower panels) obtained via stochastic reconstruction by successively incorporating higher-order $P_n$ functions. The functions incorporated from left to right are respectively: $P_2$, $\{P_2, P_3\}$, $\{P_2, P_3, P_4\}$, $\{P_2, P_3, P_4, P_6\}$, and $\{P_2, P_3, P_4, P_6, P_8\}$.

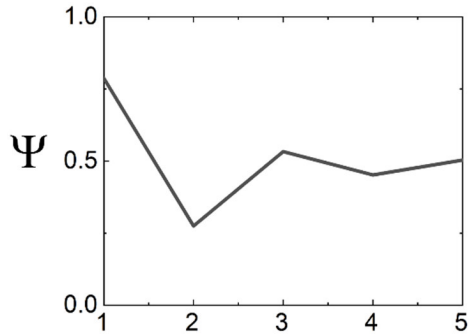

**Fig. 17:** Accuracy metric $\psi$ defined in Eq. (4) associated with reconstructions incorporating different sets of $P_n$ functions. The horizontal axis indicates the number of functions incorporated in the reconstruction.

Fig. 16 shows the realizations (upper panels) obtained via stochastic reconstruction by successively incorporating higher-order $P_n$ functions. The functions incorporated from left to right are respectively: $P_2$, $\{P_2, P_3\}$, $\{P_2, P_3, P_4\}$, $\{P_2, P_3, P_4, P_6\}$, and $\{P_2, P_3, P_4, P_6, P_8\}$. In all the reconstructions, the morphology and connectivity of both the ceramic and metallic phases have been very well reproduced. As can be seen from the comparison of the lineal-path functions (Fig. 16, lower panels) and the accuracy metric $\psi$ (see Fig. 17), including higher-order $P_n$ functions leads to successively improved reconstructions. This indicates that the additional morphological information encoded in higher-order $P_n$ functions has been efficiently utilized in the reconstructions. We note that it is relatively easier for the reconstruction algorithm to converge to a realization with connected phase at high volume fractions, as such realizations are more degenerate [94, 95]. Therefore, it is common that better reconstruction accuracy can be achieved for systems where the phase of interest (i.e., the ceramic phase) is connected.

## 4. Conclusions and Discussion

In this work, we employ stochastic realization reconstruction to probe the level of statistical morphological information contained in a recently introduced set of hierarchical statistical microstructural descriptors, i.e., the "*n*-point polytope functions" $P_n$. The $P_n(r)$ function provides the probability of finding a set of n points sitting at the vertices a regular n-polytope of edge length *r* in the phase of interest, and thus, is a subset of the corresponding standard n-point correlation function $S_n$. In particular, $P_n$ functions up to $n = 8$ were successively incorporated into the Yeong-Torquato reconstruction procedure and the accuracy of the reconstructed systems was quantitatively assessing via the lineal-path function, which provides "linear clustering" information. We examined a wide spectrum of representative random systems with distinct geometrical and topological features, including representative crystalline and disordered particle particles, Poisson distribution of particles, microstructures of concrete and interpenetrating metal-ceramic composites.

We found that generally, successively incorporating higher order $P_n$ functions which encodes essential higher-order morphological information leads to superior accuracy of the reconstructions. However, incorporating more $P_n$ functions into the reconstruction also significantly increases the complexity and roughness of the associated energy landscape for the underlying stochastic optimization, making it difficult to convergence numerically. Another observation is that the $P_n$ functions (up to n = 8 studied here) appear to be not sensitive to topological connectedness information. This was evidenced by the significant overestimation of the clustering in the reconstruction of disordered packings of hard spheres near percolation point. These examples indicate that for certain complex systems, successively incorporating higher order correlation functions in a linear fashion might not be the best practice, as one can significantly increase the computational cost without incorporating too much useful additional morphological information. An alternative approach in such cases is to utilize non-conventional functions (e.g., those encoding clustering information) or to employ machine learning techniques to identify and extract crucial higher order correlations that leapfrog unnecessary computation of all the $P_n$ functions [109-111].

**Acknowledgement**

This work is supported by ACS Petroleum Research Fund under Grant No. 56474-DNI10 (Program manager: Dr. Burtrand Lee).

56. Tahmasebi, P. and M. Sahimi, *Cross-correlation function for accurate reconstruction of heterogeneous media.* Physical review letters, 2013. **110**(7): p. 078002.
57. Xu, H., M.S. Greene, H. Deng, D. Dikin, C. Brinson, W.K. Liu, C. Burkhart, G. Papakonstantopoulos, M. Poldneff, and W. Chen, *Stochastic reassembly strategy for managing information complexity in heterogeneous materials analysis and design.* Journal of Mechanical Design, 2013. **135**(10): p. 101010.
58. Gerke, K.M., M.V. Karsanina, R.V. Vasilyev, and D. Mallants, *Improving pattern reconstruction using directional correlation functions.* EPL (Europhysics Letters), 2014. **106**(6): p. 66002.
59. Xu, H., Y. Li, C. Brinson, and W. Chen, *A Descriptor-Based Design Methodology for Developing Heterogeneous Microstructural Materials System.* Journal of Mechanical Design, 2014. **136**(5): p. 051007.
60. Gerke, K.M. and M.V. Karsanina, *Improving stochastic reconstructions by weighting correlation functions in an objective function.* EPL (Europhysics Letters), 2015. **111**(5): p. 56002.
61. Liu, X. and V. Shapiro, *Random heterogeneous materials via texture synthesis.* Computational Materials Science, 2015. **99**: p. 177-189.
62. Bostanabad, R., A.T. Bui, W. Xie, D.W. Apley, and W. Chen, *Stochastic microstructure characterization and reconstruction via supervised learning.* Acta Materialia, 2016. **103**: p. 89-102.
63. Turner, D.M. and S.R. Kalidindi, *Statistical construction of 3-D microstructures from 2-D exemplars collected on oblique sections.* Acta Materialia, 2016. **102**: p. 136-148.
64. Karsanina, M.V. and K.M. Gerke, *Hierarchical Optimization: Fast and Robust Multiscale Stochastic Reconstructions with Rescaled Correlation Functions.* Physical review letters, 2018. **121**(26): p. 265501.
65. Feng, J., Q. Teng, X. He, and X. Wu, *Accelerating multi-point statistics reconstruction method for porous media via deep learning.* Acta Materialia, 2018. **159**: p. 296-308.
66. Li, H., N. Chawla, and Y. Jiao, *Reconstruction of heterogeneous materials via stochastic optimization of limited-angle X-ray tomographic projections.* Scripta Materialia, 2014. **86**: p. 48-51.
67. Li, H., S. Kaira, J. Mertens, N. Chawla, and Y. Jiao, *Accurate Stochastic Reconstruction of Heterogeneous Microstructures by Limited X-ray Tomographic Projections.* Journal of microscopy, 2016. **264**: p. 339.
68. Li, H., S. Singh, S. Kaira, J. Mertens, J.J. Williams, N. Chawla, and Y. Jiao, *Microstructural Quantification and Property Prediction Using Limited X-ray Tomography Data.* JOM, 2016. **68**: p. 2288.
69. Kamrava, S., M. Sahimi, and P. Tahmasebi, *Quantifying accuracy of stochastic methods of reconstructing complex materials by deep learning.* Physical Review E, 2020. **101**(4): p. 043301.
70. Kalidindi, S.R., *Hierarchical Materials Informatics: Novel Analytics for Materials Data.* 2015: Elsevier.
71. Kalidindi, S.R., J.A. Gomberg, Z.T. Trautt, and C.A. Becker, *Application of data science tools to quantify and distinguish between structures and models in molecular dynamics datasets.* Nanotechnology, 2015. **26**(34): p. 344006.
72. Steinmetz, P., Y.C. Yabansu, J. Hötzer, M. Jainta, B. Nestler, and S.R. Kalidindi, *Analytics for microstructure datasets produced by phase-field simulations.* Acta Materialia, 2016. **103**: p. 192-203.
73. Fullwood, D.T., S.R. Niezgoda, B.L. Adams, and S.R. Kalidindi, *Microstructure sensitive design for performance optimization.* Progress in Materials Science, 2010. **55**(6): p. 477-562.
74. Jain, A., J.R. Errington, and T.M. Truskett, *Dimensionality and design of isotropic interactions that stabilize honeycomb, square, simple cubic, and diamond lattices.* Physical Review X, 2014. **4**(3): p. 031049.
21